# Fault–based recurrence models and occurrence probabilities of large earthquakes (M≥6.0) in the Corinth Rift, Greece

Kourouklas C.[1], Bonatis P.[1], Papadimitriou E.[1], Karakostas V.[1]

(1) Geophysics Department, School of Geology, Aristotle University of Thessaloniki, Thessaloniki, Greece, ckouroukl@geo.auth.gr.

## Research Highlights

- The $M≥6.0$ earthquakes recurrence times, $T_r$, exhibit high variability from 40 to 1500 years, with the southern Corinth Rift fault segments reaching values up to 350 years and their antithetic ones ranging from 400 to 1500 years.
- The fault segments in the Corinth Rift can be divided into three groups, according to their recurrence behaviour, with some of them exhibiting significantly lower renewal model probabilities than the Poisson model, others showing nearly equal probabilities, and one segment where the renewal model probabilities are much higher than the Poisson one.

## Introduction

The main objectives of the current study include the determination of the mean recurrence time, $T_r$, of large ($M≥6.0$) earthquakes associated with the major fault segments of the Corinth Rift, Greece, and the estimation of the conditional occurrence probabilities of an impending $M≥6.0$ earthquake on each major fault segment.

The recurrence behavior of large earthquakes (e.g. $M≥6.0$) on specific fault segments is one of the primary input parameters for developing long–term Earthquake Rupture Forecast (ERF) models. These models integrate a series of parameters (maximum observed magnitude, fault dimensions, long–term slip rates, available recurrence times) to estimate the occurrence likelihood of nearly characteristic magnitude earthquakes in a specific time span. The primary output of such models is the mean recurrence time, which can later be used for the application of statistical models, which, in turn, return the likelihood of the occurrence of near characteristic magnitude earthquakes over specific time intervals and can be based on either a time–independent or an elastic rebound motivated renewal assumption.

A precise and robust estimation of $T_r$ requires the inclusion of as many large earthquakes associated with individual fault segments as possible, including both historical and instrumental data along with the selection of the appropriate statistical model. However, large earthquakes associated with specific fault segments are often limited, with only a few cases having about 3 to 10 observations due to the long duration required for stress accumulation and the short time span of available large earthquake records.

To address these challenges and limitations, an alternative approach is the estimation of $T_r$ through the application of the seismic moment rate conservation method (Field *et al.*, 1999). This method defines the mean recurrence time as the ratio of the maximum expected seismic moment corresponding to a large earthquake with the maximum observed magnitude ($M_{max}$) occurring on a given fault segment, to the seismic moment that might be released by the respective fault segment due to the tectonic loading, assuming a nearly characteristic earthquake model. This approach can provide more precise $T_r$ estimates because the overall rate and size distribution of earthquakes should reflect the tectonic loading in the brittle part of the crust, and this function is typically constrained by the principle of seismic moment rate conservation.

## Seismotectonic Setting of Corinth Rift

The Corinth Rift, a prominent E–W oriented elongated graben stretching over 100 km in length, separates the Peloponnese from mainland Greece. It exhibits a remarkable extension rate that varies spatially, with the western section extending at approximately 10 mm/yr and the eastern section at around 5.5 mm/yr (Briole *et al.*, 2021). As a result, the Corinth Rift is one of the most seismically active regions, frequently experiencing large ($M≥6.0$) earthquakes during both the instrumental and historical seismicity periods (Papazachos & Papazachou, 2003), with the most recent one being the 1995 $M_w$=6.5 Aigion earthquake (Bernard *et al.*, 1997). The intense seismic activity along the Corinth Rift has made it one of the most well studied areas, resulting in the detailed determination of its main fault segments. Information on the $M≥6.0$ earthquakes are derived from the catalogue provided by Papazachos & Papazachou (2003) and regional parametric earthquake catalog of the Geophysics Department of the Aristotle University of Thessaloniki, consisting of 35 large earthquakes since 373 BC. Studies by Console *et al.* (2013, 2015), after concluding that the large earthquakes catalog could be considered as complete after 1700 AD (since 1714 AD) as well as Kourouklas (2022), suggest that they are not regularly distributed but they are clustered in time. This latter dataset includes 22 $M\geq6.0$ earthquakes (Figure 1).

The southern margin of the Corinth Rift is bounded by eight (8) normal faults, listed from west to east as Psathopyrgos (FS 01), Aigion (FS 02), Eliki (FS 03), Akrata (FS 04), Xylokastro (FS 05), Perachora (FS 06), Skinos (FS 07) and Alepochori (FS 08) fault segments (Figure 1). Their strike values are almost the same 260° – 280° and their dip angles vary between 30° and 45°. The northeastern boundary of study area is demarcated by the Kapareli fault segment (FS 09), also a normal fault with a strike of 50° and a dip angle of 50°. The fault segment lengths range from 8 km (Akrata Fault Segment) to 22 km (Eliki Fault Segment). All their geometrical and kinematic were summarized by Console *et al.* (2013, and references therein). To the north the Delfoi (FS 10), Makrygialos (FS 11), Sykia (FS 12) and Marathias (FS 13) fault segments are defined from east to west (Figure 1; Ganas 2024, and references therein). Their lengths range from 10 km (Makrygialos and Sykia fault segments) to 18 km (Delfoi fault segment).

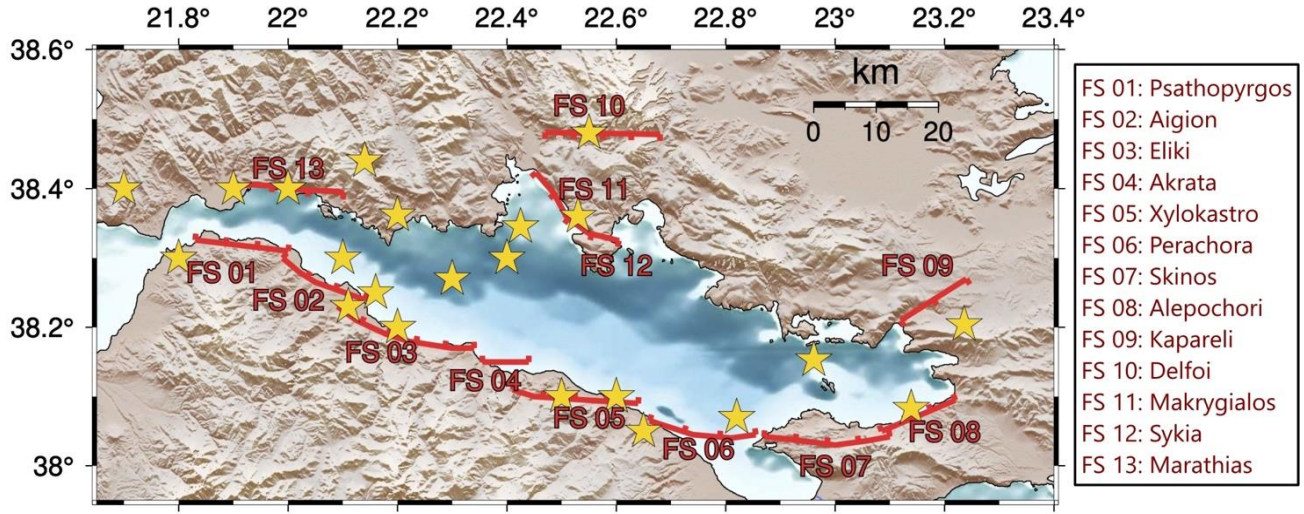

**Figure 1. Geomorphological map of Corinth Rift, where the major normal fault segments are represented by red solid lines. Yellow stars depict the $M\geq6.0$ earthquakes occurred in the study area since 1714.**

Although the number of large earthquakes (both historical and instrumental) is adequate for regional statistical studies, the number of recurrence intervals become limited when subdivided into subsets associated with specific fault segments. For this reason, Kourouklas (2022, and references therein) attempted to associate the 22 $M\geq6.0$ earthquakes since 1714 with their respective causative fault segments. The analysis suggested that 2 out of 8 fault segments of the western Corinth Rift were associated with two (2) $M\geq6.0$ earthquakes each (Psathopyrgos and Eliki). The Aigion fault segment was linked to four (4) events, whereas the Xylokastro and Perachora segments were each associated with three (3). The Alepochori and Skinos fault segments were associated with one (1) $M\geq6.0$ earthquake each. It is noteworthy that the Akrata fault segment, the shortest in length, has not been associated with any $M\geq6.0$ earthquake since 1714. Similarly, the Kapareli fault segment was also associated with only one (1) $M\geq6.0$ earthquake. In the northern part, the Delfoi, Makrygialos and Sykia segments were associated with one (1) $M\geq6.0$ historical earthquake each, whereas the Marathias fault was linked to two (2) such events. Thus, although the study area is among the most seismically active areas where several large earthquakes rather frequently take place, the insufficient number of $M\geq6.0$ earthquake events per segment restricts the ability to determine $T_r$ by using their observational recurrence intervals as input.

**Method**

We estimated the mean recurrence time, $T_r$, by applying the seismic moment rate conservation technique (Field *et al.*, 1999) in the absence of a sufficient number of recurrence intervals for each of the 13 normal faults in the Corinth Rift. This technique assumes that a large earthquake of nearly characteristic magnitude can release the total accumulated seismic moment on a fault. It takes into account the maximum observed magnitude ($M_{max\_obs}$) and its corresponding uncertainty ($\Delta M$) and the maximum seismic moment corresponding to the accumulated strain on the specific fault segment due to tectonic loading. The mean recurrence time is then calculated as the ratio of the seismic moment rate that can be released by the maximum magnitude earthquake to the seismic moment due to strain accumulation on the fault:

$$T_r = \frac{M_{o\,max}}{\mu L w V} \quad (1)$$

where $M_{o,max}$ is the maximum possible seismic moment released by a large earthquake with a magnitude within the range $M_{max\_obs} \pm \Delta M$. $\mu$ is the shear modulus, whose typical value for faults in the Earth's crust is equal to $3.3 \times 10^5$ bar ($\mu = 3.3 \times 10^5$ bar), $L$ and $w$ are the length and width of the fault segment (in km), respectively, and $V$ the long–term slip rate.

The maximum expected seismic moment is computed via the definition of seismic moment. The variability of $T_r$ is also computed via the method of formal error propagation technique proposed by Peruzza et al. (2010), considering the uncertainties related to the maximum magnitude and slip rate values. Knowing the variability of mean recurrence time, the standard deviation, $\sigma$, can then be easily calculated. The advantage of this approach is that, once $T_r$ and $\sigma$ are estimated, the corresponding aperiodicity values, $\alpha$, can be calculated as the coefficient of variation, $C_v$, of $T_r$, which is the second appropriate parameter of renewal model applications (Pace et al., 2016).

Using the estimated $T_r$ and $\sigma$ as the primary inputs, we model the $M \geq 6.0$ earthquakes recurrence times by applying both the time–independent Poisson model and the Brownian Passage Time (BPT; Matthews et al., 2002) renewal model, aiming at the estimation of the occurrence probabilities of the next $M \geq 6.0$ earthquake to be occurred on each fault segment for the next 10, 20 and 30 years, starting from 1 January 2025.

The Poisson process can be expressed by the exponential distribution with probability density function (pdf) given by:

$$f(t|T_r) = \frac{1}{T_r} \exp\left\{-\frac{t}{T_r}\right\} \quad (2)$$

where $T_r$ is the mean recurrence time of large earthquakes associated with a specific fault segment. To model the occurrence of large earthquakes as a renewal process, the BPT distribution is applied. The PDF of the BPT model is given by:

$$f(t|T_r, \alpha) = \left(\frac{T_r}{2\pi \alpha^2 t^3}\right)^{1/2} \exp\left\{-\frac{(t-T_r)^2}{2T_r \alpha t}\right\} \quad (3)$$

where $T_r$ is also the mean recurrence time and $\alpha$ is the aperiodicity, which can be considered as analogous to the coefficient of variation, $C_v$, of the normal distribution.

The occurrence probabilities of a next large earthquake on certain fault in a specific time span can be computed by applying the simple Poisson probability model given by:

$$P(t \leq T \leq t + \delta t) = 1 - e^{-\delta t/T_r} \quad (4)$$

where $\delta t$ is the forecast duration and the conditional probability corresponding to the BPT renewal given by:

$$P(t \leq T \leq t + \delta t) = \frac{\int_t^{t+\delta t} f(t)dt}{\int_t^{\infty} f(t)dt} \quad (5)$$

where $t$ is relative to the date of last earthquake, conditioned by the fact that it has been $t + \delta t$ years since the last one.

Alternatively, the occurrence probabilities of the next large earthquake can be evaluated by the estimation of the hazard function, $H(t)$, of both Exponential and BPT models. This analysis is particularly useful for predicting future rupture scenarios as the values of the hazard function or, in other words, the hazard rate is equivalent to the conditional probability estimate within a specific time span. The hazard function of a given distribution can be easily defined using its corresponding probability density, $f(t)$, and cumulative density, $F(t)$, functions as follows:

$$H(t) = \frac{f(t)}{S(t)} = \frac{f(t)}{1-F(t)} \quad (6)$$

where $S(t)$ is the survival function of the distribution.

**Application & Results**

The mean recurrence times, $T_r$, of $M \geq 6.0$ earthquakes are estimated through Equation (1) after defining all the appropriate parameters for the application of the seismic moment rate conservation method. These parameters include the fault dimensions ($L$, $w$), the maximum observed magnitude, $M_{\_max\_obs}$, their long-term slip rates ($V$) and the corresponding uncertainties ($\Delta M$ and $\sigma V$, respectively) which are summarized in Kourouklas (2022). It should be noted that for the definition of the long-term slip rates assigned to each fault segment, 60% of the geodetic slip rates proposed by Briole et al. (2021) are used. As previously mentioned, the Akrata fault segment was not associated with any $M \geq 6.0$ earthquake and the estimation of the mean recurrence time and the maximum expected magnitude, were calculated using scaling relations, The absence of any $M \geq 6.0$ earthquake associated with this fault segment prevents the estimation of any elapsed time, which is necessary for renewal recurrence modeling. For this reason, the Akrata fault segment is excluded from the next steps of the statistical analysis.

The estimated $T_r$ values (Figure 2a) range from a few years (Akrata fault segment; $T_r$=35 years) to about 1500 years (Makrygialos fault segment; $T_r$=1511.5 years). These variations are clearly related to the dimensions and slip rates of each segment, which results in different stressing rate values and maximum observed magnitudes. Starting from the southern Corinth Rift fault zone (excluding the Akrata fault segment), mean recurrence times vary from almost 120 years (Psathopyrgos fault segment; $T_r$=119.4 years) to 359 years (Skinos fault segment; $T_r$=359.3 years) (Figure 2a). The $T_r$ of $M \geq 6.0$ earthquakes increase from the western part to the eastern one (Figure 2a), due to the gradual decrease of deformation rates from the western to eastern part of our study area. Regarding the respective $C_v$ values, it is reported that they exhibit minimal variations, ranging from 0.6 to 0.7 in 6 out of 7 cases (Psathopyrgos, Aigion, Xylokastro, Perachora, Skinos, Alepochori). This suggests that the recurrence behavior of $M \geq 6.0$ earthquakes associated with these fault segments can be characterized as slightly aperiodic. On the other hand, results of our analysis for $M \geq 6.0$ earthquakes associated with the Eliki fault segment indicates that its recurrence behavior can be characterized as highly aperiodic (almost random), given that its $C_v$ value is equal to 0.9 ($C_v$=0.9).

Moving to the mean recurrence time results for the northern part, higher values are reported, ranging from 426 years to 1511 years. Specifically, the $T_r$ of $M≥6.0$ earthquakes associated with the Kapareli fault segment is found equal to 426 years, with a $Cv$ value of 0.6. The mean recurrence time for the Delfoi fault segment is equal to $T_r$=933.1 years, whereas the $T_r$ of $M≥6.0$ earthquakes associated with Makrygialos and Sykia fault segments are even larger, with $T_r$=1511.5 years and $T_r$=1070.1 years, respectively. The mean recurrence time of $M≥6.0$ earthquakes associated with the Marathias fault segment is $T_r$=546.1 years. The coefficient of variation for all four these cases (Delfoi, Makrygialos, Sykia, Marathias) is estimated equal to $Cv$=0.7.

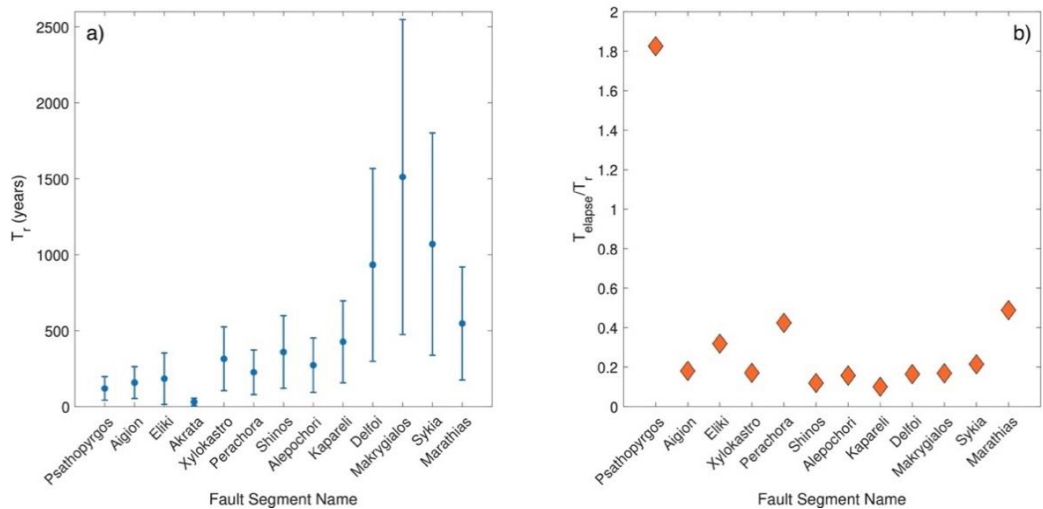

Figure 2. (a) Mean recurrence time, $T_r$, estimates of $M≥6.0$ earthquakes (blue circles), along with their ±1σ (vertical blue solid line) for the 13 normal fault segments of the Corinth Rift. (b) The ratio of elapsed time, $T_e$, of the last $M≥6.0$ earthquake that occurred in each fault segment (except from Akrata fault segment) and the mean recurrence time, $T_r$, estimates.

One more important factor for the statistical modeling of $T_r$, especially for time–dependent approaches, is the elapsed time, $T_e$, since the last earthquake. Specifically, the ratio of $T_e$ over $T_r$ can serve as an initial indicator of the stage of the earthquake cycle for a given fault. Values of this ratio approaching zero indicate the early stage of a new earthquake cycle, while more increased values indicate that the fault is closer to the next rupture. Figure 2b shows the ratio of the elapsed time to the mean recurrence time for the seven fault segments. The smallest value is reported for the Aigion fault segment, which is the most recently ruptured ($T_e$ = 28.57 years). For 10 out of the other 11 fault segments, the ratio ranges between 0.17 and 0.55, also showing that the elapsed time is considerably shorter than the mean recurrence time. This suggests that these faults are likely in the early stages of their seismic cycle. The only case in which the ratio is larger than 1.0 is the Psathopyrgos fault segment (Figure 2b), meaning that the elapsed time since the last $M≥6.0$ is much larger than the estimated $T_r$.

Next, we applied the exponential and BPT distributions, representing the Poisson and renewal models, using the mean recurrence time, $T_r$, and $Cv$ values obtained from the seismic moment rate conservation approach. The ultimate goal is to model the recurrence behavior of the $M≥6.0$ earthquakes associated with the normal faults of the study area. We focused primarily on the hazard functions of both statistical models, as their (hazard) rates can be considered equivalent to the conditional probability estimates over a specific time span. Figure 3 shows the hazard functions of the exponential and BPT distributions for the seven (7) fault segments of the southern part (Psathopyrgos, Aigion, Eliki, Xylokastro, Perachora, Skinos and Alepochori), in which their significant differences in modeling $T_r$ are highlighted.

Starting from the Psathopyrgos fault segment (Figure 3a) the constant hazard rate of the Poisson model (blue solid line) is considerably lower than the hazard rate of the BPT model (red solid line), which exhibits a decreasing trend, at the time corresponding to $T_e$ (until 31-12-2024). On the other hand, in the case of the Aigion fault segment (Figure 3b), the most recently ruptured segment in the study area, the hazard function values of the Poisson model (blue solid line) are considerably larger than the values obtained from the application of the BPT, at the time corresponding to $T_e$. The BPT model application on the Eliki fault segment (Figure 3c) reports hazard rates slightly higher than those of the Poisson model around the time corresponding to the elapsed time (black dashed line in Figure 3c). More specifically, the BPT hazard function values exhibit an increasing trend for the next 10, 20 and 30 years after 31-12-2024. Similarly, for the Aigion fault segment, the hazard function values of the Poisson model (blue solid line) are considerably larger than the values of the BPT model for the recurrence behavior of the $M≥6.0$ earthquakes associated with the Xylokastro fault segment (Figure 3d), indicating that the fault segment is at an early stage of a new earthquake cycle at time $t$ corresponding to the elapsed time. For the Perachora fault segment both models report almost equal hazard rates (Figure 3e) and lastly, for the Skinos and Alepochori fault segments (Figures 3f and 3g, respectively) the hazard function values of the BPT model are significantly lower than the ones of the Poisson model at time $t$ corresponding to the elapsed time, highlighting that both segments are at an early stage of a new earthquake cycle according to the renewal model.

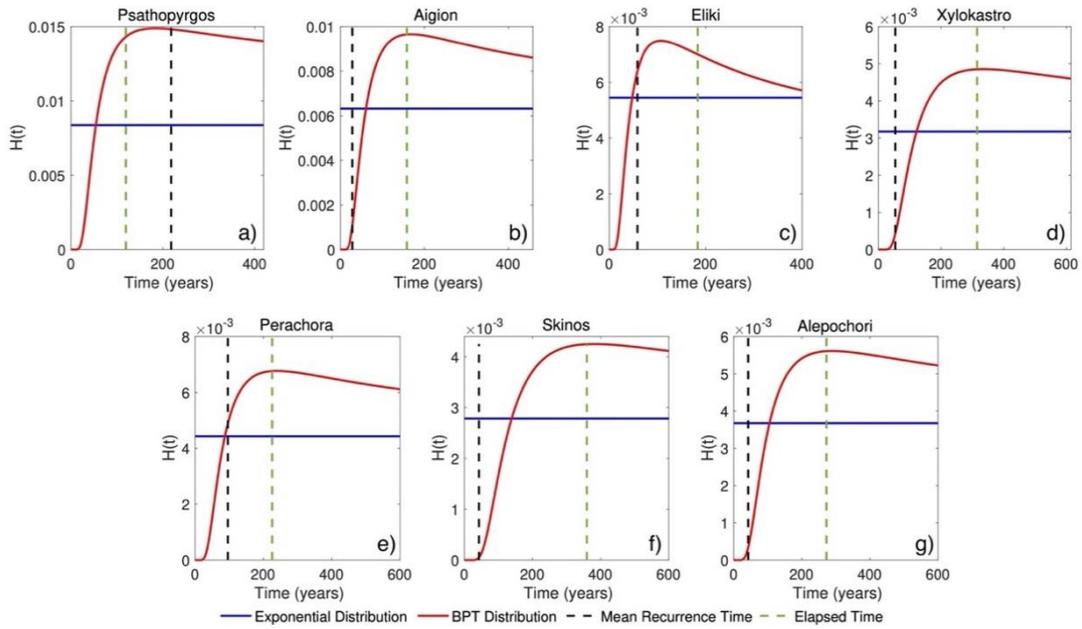

**Figure 3.** Hazard functions, *H(t)*, for the Psathopyrgos (a), Aigion (b), Eliki (c), Xylokastro (d), Perachora (e), Skinos (f) and Alepochori (g) fault segments according to the Exponential (blue solid lines) and the BPT (solid red lines) distributions. Vertical dashed lines denote the elapsed time since the last *M*≥6.0 earthquake (*t* = 0) occurred on each fault segment, whereas the vertical dashed black lines denote the mean recurrence time of *M*≥6.0 earthquake for each fault segment.

Focusing on the recurrence modeling results for the five (5) fault segments of the northern Corinth Rift fault zone, in 4 out of 5 cases the hazard rate values of the renewal model are significantly lower than those of the Poisson model. Specifically, the statistical modeling of *M*≥6.0 earthquakes associated with the Kapareli (Figure 4a), Delfoi (Figure 4b), Makrygialos (Figure 4c) and Sykia (Figure 4d) fault segments show that the hazard rate estimates of the time–independent Poisson model are considerably higher (blue solid lines in Figures 4a, 4b, 4c and 4d) than the ones of the BPT model (solid red lines). More specifically, for the Kapareli fault segment, the BPT model hazard function values are almost zero at the time corresponding to the elapsed time, whereas the constant hazard rate of the Poisson mode is equal to $2.4 \times 10^{-3}$. Statistical modeling for Delfoi, Makrygialos and Sykia fault segments indicates slightly different results for the renewal model, with its hazard function values exhibiting an increasing trend for the next 10, 20 and 30 years after elapsed time (vertical dashed black lines in Figures 4b, 4c and 4d) in all cases. On the contrary, the recurrence modeling of *M*≥6.0 earthquakes associated with the Marathias fault segment (Figure 4e) reports hazard rates slightly higher than the ones of the Poisson model around the time corresponding to the elapsed time.

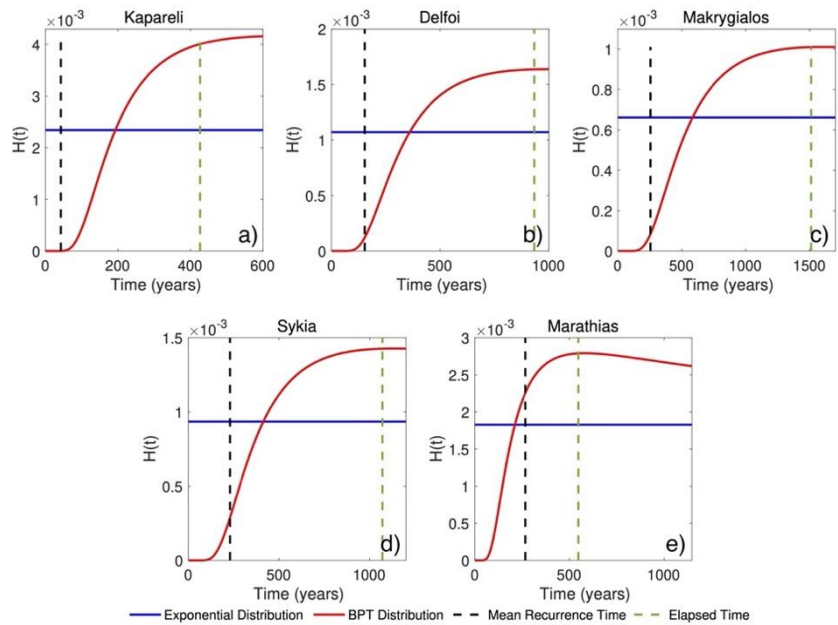

**Figure 4.** Hazard functions, *H(t)*, for the Kapareli (a), Delfoi (b), Makrygialos (c), Sykia (d) and Marathias (e) fault segments according to the Exponential (blue solid lines) and the BPT (solid red lines) distributions. The vertical dashed lines denote the elapsed time since the last *M*≥6.0 earthquake (*t* = 0) occurred on each fault segment, whereas the vertical dashed black lines denote the mean recurrence time of *M*≥6.0 earthquake for each fault segment.

## Conclusions

In this study, we present both time–dependent and time–independent recurrence models to investigate the mean recurrence time of $M≥6.0$ earthquakes associated with the main normal faults of the Corinth Rift. We used the most detailed available input data (fault network model, long–term slip rates) for the estimation of the mean recurrence time, $T_r$, by applying the physics–based seismic moment rate conservation method. The results show that $T_r$ ranges from about 40 years to almost 1500 years, depending on the combination of the input parameters. The statistical analysis of $T_r$ via the Exponential and BPT distributions (representing the Poisson and a renewal model, respectively) reveals that the recurrence behavior of the Corinth Rift fault segments can be divided into three distinct groups. The first group includes fault segments where the elapsed time since the last earthquake is considerable smaller than the $T_r$. This group includes the Aigion, Xylokastro, Skinos, Alepochori, Kapareli, Delfoi, Makrygialos and Sykia fault segments. For all these cases the renewal model's hazard function values and the corresponding conditional occurrence probabilities for the next 10, 20 and 30 years show significant differences among the two applied models, with the BPT model yielding much lower values than the Poisson model. These findings indicate that the aforementioned fault segments are at an early stage of a new earthquake cycle according to the renewal model. The second group includes the Eliki, Perachora and Marathias fault segments for which the hazard rates of both models are almost equal, resulting in similar occurrence probabilities for the next 10, 20 and 30 years. The last group includes only the Psathopyrgos fault segment for which the BPT model reports significantly larger hazard rate values than the Poisson model. The Psathopyrgos fault segment is the only case where the elapsed time since the last $M≥6.0$ earthquake exceeds the mean recurrence time, indicating that from a statistical point of view, it is at a later stage of a new earthquake cycle. Consequently, the occurrence probabilities for the next $M≥6.0$ earthquake to occur in the next 10, 20 and 30 years are the highest within the study area. The results of our analysis could serve as the basis for fault–based large earthquake occurrence models. By considering time–dependent occurrence models in probabilistic seismic hazard analysis (PSHA) one can account for epistemic uncertainties, quantify the range of seismic hazard for a given exceedance probability and illustrate which model assumptions lead to the most significant variations in the estimated seismic hazard (Akinci *et al.*, 2017). By utilizing these models, stakeholders can better determine when and where protective measures are most urgently needed, particularly in regions with higher short-term earthquake probabilities, leading to more effective seismic risk management.

## Acknowledgements

This research is financially supported by the artEmis Project funded by the European Union, under Grant Agreement nr 101061712. Views and opinions expressed are however those of the author(s) only and do not necessarily reflect those of the European Union or European Commission – Euratom. Neither the European Union nor the granting authority can be held responsible for them.

## References

Akinci, A., Vannoli, P., Falcone, G., Taroni, M., Tiberti, M.M., Murru, M., Burrato, P., Mariucci, M.T., 2017. When time and faults matters: Towards a time-dependent probabilistic SHA in Calabria, Italy. Bulletin of Earthquake Engineering 15, 2497-2524.

Bernard, P., Briole, P., Meyer, B., Lyon-Caen, H., Gomez, J.-M., Tiberi, C., Berge, C., Cattin, R., Hatzfeld, D., Lachet, C., Lebrun, B., Deschamps, A., Courboulex, F., Larroque, C., Rigo, A., Massonet, D., Papadimitriou, P., Kassaras, I., Diagourtas, D., Makropoulos, K., Veis, G., Papazisi, E., Mitsakaki, C., Karakostas, V., Papadimitriou, E., Papanastassiou, D., Chouliaras, G., Stavrakakis, G., 1997. The Ms = 6.2, June 15, 1995 Aigion earthquake (Greece): Evidence for low angle normal faulting in the Corinth rift. Journal of Seismology 1, 131–150. Doi:https://doi.org/10.1023/A:1009795618839.

Briole, P., Ganas, A., Elias, P., Dimitrov, D., 2021. The GPS velocity field of Aegean. New observations, contribution of the earthquakes, crustal blocks model, Geophysical Journal International 226, 468-492.

Console, R., Falcone, G., Karakostas, V., Murru, M., Papadimitriou, E., Rhoades, D., 2013. Renewal models and coseismic stress transfer in the Corinth Gulf, Greece, fault system/ Journal of Geophysical Research, Solid Earth 118, 3655–3673. Doi:https://doi.org/10.1002/jgrb.50277.

Console, R., Carluccio, R., Papadimitriou, E., Karakostas, V., 2015. Synthetic earthquake catalogs simulating seismic activity in the Corinth Gulf, Greece, fault system. Journal of Geophysical Research 120, 326–343. Doi:https://doi.org.10.1002/2014JB011765.

Field, E.H., Jackson, D.D., Dolan, J.F., 1999. A mutually consistent seismic hazard source model for Southern California. Bulletin of Seismological Society of America 89, 559–578.

Ganas, A., 2024. NOAFAULTS KMZ layer Version 6.0 (version 6.0) [Data set]. Zenodo. Doi: https://doi.org/10.5281/zenodo.13168947

Kourouklas, C., 2022. Determination and simulation of strong earthquakes' recurrence times in Greece via the application of stochastic models: contribution on seismic hazard assessment. Ph.D. Thesis, Aristotle University of Thessaloniki, Thessaloniki, 329 p. Doi: http://dx.doi.org/10.12681/eadd/53270

Matthews, M.V., Ellsworth, W.L., Reasenberg, P.A., 2002. A Brownian model for recurrent earthquakes, Bulletin of Seismological Society of America 92, 2233–2250.

Pace, B., Visini, F., Peruzza, L., (2016). FiSH: MATLAB Tools to Turn Fault Data into Seismic-Hazard Models, Seismological Research Letters, 87, 375-386.

Papazachos, B.C., Papazachou, C.C., 2003. The earthquakes of Greece, Ziti Publications.

Peruzza, L., Pace, B., Cavallini, F., 2010. Error propagation in time- dependent probability of occurrence for characteristic earthquakes in Italy, Journal of Seismology, 14, 119–141.